\documentclass[preprint,prl,amssymb]{revtex4}
\usepackage{graphicx}% Include figure files
\usepackage{dcolumn}% Align table columns on decimal point
\usepackage{bm}% bold math
\usepackage{psfrag,epsfig}
\def\be{\begin{equation}}
\def\ee{\end{equation}}
\def\Be{\begin{eqnarray}}
\def\Ee{\end{eqnarray}}
\def\ba{\begin{array}}
\def\ea{\end{array}}
\begin{document}

\title{Reply to Comment on ``Mass and $K\Lambda$ Coupling of the $N^*(1535)$"}

\author{B.C.Liu and B.S.Zou}
\affiliation{Institute of High Energy Physics, P.O.Box 918(4),
Beijing 100049, China}
%\date{\today}

%\pacs{36.10.Gv, 11.80.-m, 12.39.Fe, 13.75.Jz}

\maketitle

\noindent

In the comment \cite{sibirtsev}, the authors criticize our
calculation \cite{liu} of $pp\to p K^+ \Lambda$ near threshold for
not taking into account the $\Lambda p$ final state interaction
(FSI), and consequently question our conclusions based on the
large $N^*(1535)$ coupling to the $K\Lambda$ system. It is true
that the $\Lambda p$ FSI is important for $pp\to p K^+ \Lambda$
near threshold as clearly shown by a recent COSY experiment
\cite{cosy} and should not be neglected. However, since there are
large uncertainties for other elements in calculations of $pp\to p
K^+ \Lambda$, here we will show that the large $N^*(1535)$
coupling to the $K\Lambda$ deduced from the BES data on
$J/\psi\to\bar p K^+ \Lambda$ \cite{bes2} and $J/\psi\to\bar p
p\eta$ \cite{ppeta} is still compatible with the COSY results on
$pp\to p K^+ \Lambda$ after including the $\Lambda p$ FSI.

Some ingredients with large uncertainties for calculating $pp\to p
K^+ \Lambda$ are : 1) forms and parameters of form factors for
hadronic coupling vertices; 2) parameters of resonances such as
mass, width and coupling constants; 3) interference terms between
different resonances; 4) parameters for the $\Lambda p$ FSI.
Although these ingredients need to be considered consistently for
all possible relevant processes, there is always some room for
adjustment.

In our previous calculation \cite{liu} of $pp\to p K^+ \Lambda$,
following Ref.\cite{tsushima}, we neglected the interference terms
between different resonances and the $\Lambda p$ FSI. We added the
$N^*(1535)$ contribution with $g_{N^*(1535)K\Lambda}=1.3
g_{N^*(1535)\eta N}$, $g_{\eta NN}^2/4\pi=5$ and $\Lambda_\eta =
2.0$ GeV for the corresponding form factor.

\begin{figure}[htbp]
\begin{center}
\includegraphics[height=6.5cm]{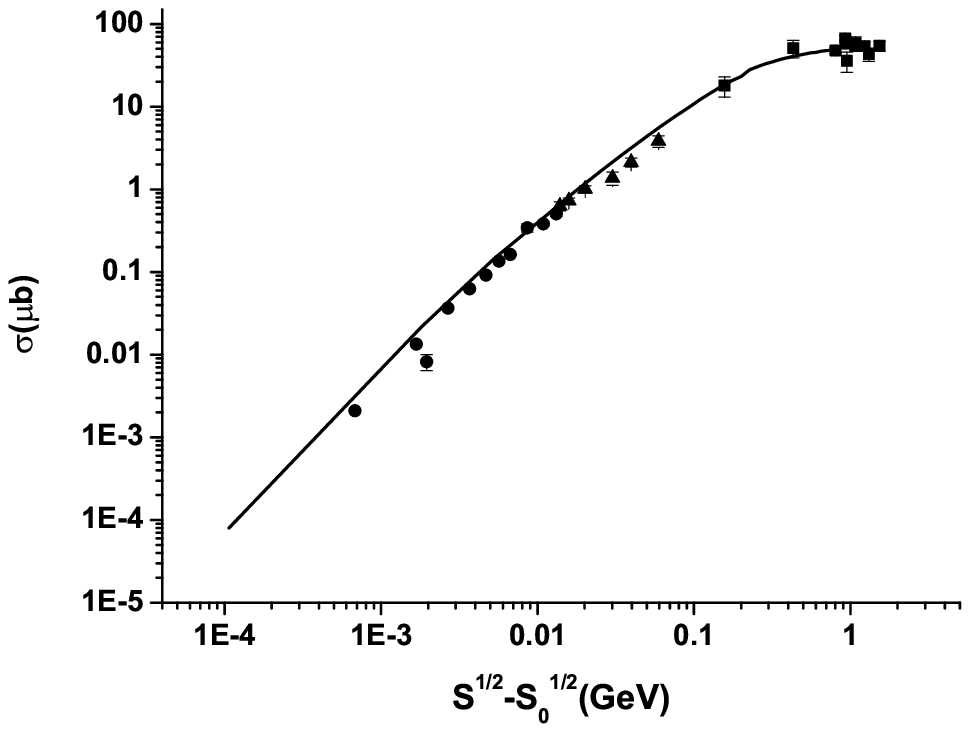}
\includegraphics[height=6.5cm]{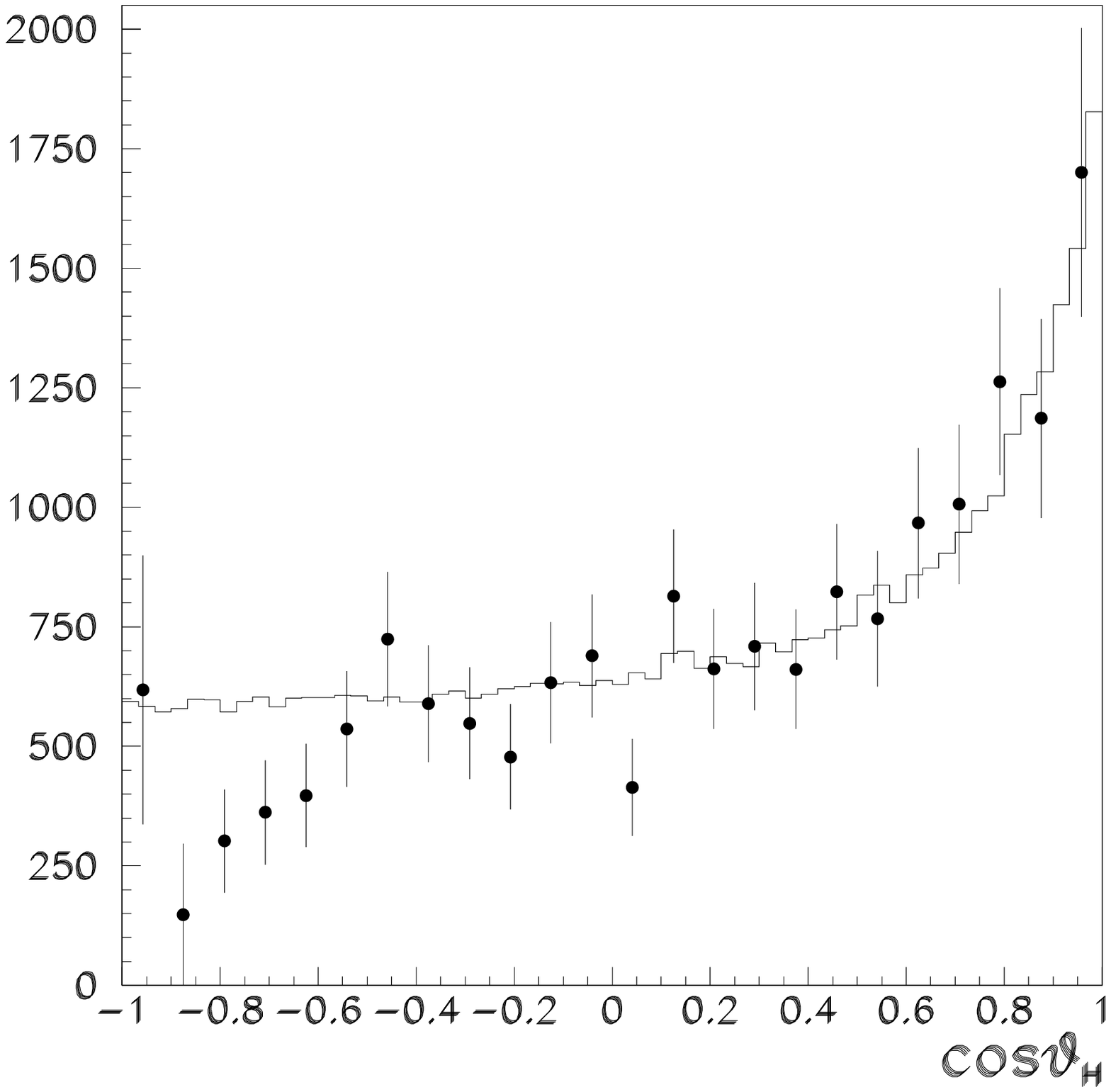}
  \caption{The total cross section vs the excess energy (left) and the $\Lambda p$
  helicity angle spectra for $K^+\Lambda$ masses $M_{K\Lambda}>1.74GeV$ at $p_{beam}$=2.8 GeV (right)
  for $pp\to p K^+\Lambda$.  }
 \label{result}
\end{center}
\end{figure}

Now in order to incorporate the $\Lambda p$ FSI effect, we just
need to adjust these parameters within their uncertainties. In the
modified calculation, we use
$g_{N^*(1535)K\Lambda}=g_{N^*(1535)\eta N}$ which is allowed by
uncertainty of BES data \cite{liu}, $g_{\eta NN}^2/4\pi=3$ and
$\Lambda_\eta=1.5$ GeV, which are well within the uncertainties by
relevant processes \cite{zhu}; for the $\Lambda p$ FSI we use the
same approach as in Ref.\cite{sibirtsev2} by adopting the
parameters $\beta =201.7$ MeV and $\alpha =-76.8$ MeV ($a= -1.59$
fm and $r= 3.16$ fm). Here we also include the the contribution
from $\rho$ meson exchange, which only gives significant
contribution at higher energy. Without changing any other
parameters, the results are shown in Fig.\ref{result}, which
already produce both total cross sections and the $\Lambda p$
helicity angle spectra quite well. Note we have not used the
freedom of adjusting more parameters, allowing free interference
terms and including the initial state interaction \cite{ISI} yet.
It just serves as an example to show that the large coupling of
$N^*(1535)$ to $K\Lambda$ deduced from BES results is still
compatible to the $pp\to p K^+ \Lambda$ experiment results within
the uncertainties after including the $\Lambda p$ FSI. In fact, by
comparing the Dalitz plots shown by Fig.5(a) and Fig.5(b) in
Ref.\cite{cosy}, it is obvious that the calculation with the
adjusted model of Sibirtsev without including the contribution of
the $N^*(1535)$ underestimates the part near $K\Lambda$ threshold.
The inclusion of the $N^*(1535)$ may reduce the $N^*(1650)$
contribution necessary to reproduce the data.

In the reaction $J/\psi\to\bar p K \Lambda $, the $K\Lambda$
invariant mass spectrum divided by the phase space factor shows a
clear peak at the $K\Lambda$ threshold without peak at 1650 MeV
(see Fig.9(b) in Ref.\cite{bes2}), so the peak is most likely due
to the sub-threshold $N^*(1535)$ resonance. For the solution using
resonance mass of 1650 MeV without including the $N^*(1535)$, the
fit is worse meanwhile with the fitted width and $K\Lambda$
branching ratio well out of the PDG range \cite{bes2}.

The $N^*(1535)$ is the most outstanding signal in both
$J/\psi\to\bar p K^+ \Lambda$ \cite{bes2} and $J/\psi\to\bar p
p\eta$ \cite{ppeta}, and is produced back-to-back against $\bar p$
with large relative momenta without the complication caused by
t-channel exchange of various mesons as in $pp\to p K^+ \Lambda$.
Hence the ratio between its decay branching ratios to $K\Lambda$
and $\eta N$ can be determined \cite{liu} more reliably than other
processes.

\end{document}